# Education and Research in the SEENET-MTP Regional Framework for Higher Education in Physics

R. Constantinescu[a] and G. S. Djordjevic[b]

[a]Department of Theoretical Physics, University of Craiova, 13 A. I. Cuza Str., 200585 Craiova Romania

[b]Department of Physics, University of Nis, P.O.Box 224, Nis, Serbia

**Abstract.** Southeastern European countries undergo significant changes in the demand/supply ratio on the labour market and in the structure of professional competences that are necessary for undertaking a professional activity. In addition, brain-drain process and decrease of interest for a career in basic sciences put many challenges for our community. Consequently, based on the activity of the Southeastern European Network in Mathematical and Theoretical Physics (SEENET MTP Network) in connecting groups and persons working in mathematics and theoretical physics, we investigate specific qualifications recognized in these fields in all the countries from the region, and the related competences necessary for practising the respective occupations. A list of new possible occupations will be promoted for inclusion in the National Qualifications Register for Higher Education. Finally, we analyze the vision existing in this region on the higher education qualifications against the European vision and experience, in particular in training of Master students, PhD students, and senior teaching and research staff through the Network, i.e. multilateral and bilateral programs.

**Keywords:** research and education, networking,

## INTRODUCTION

Recognizing the importance, as well as the necessity, of bridging the gap between Southeastern and Western European scientific community, the participants of the UNESCO-ROSTE-sponsored BALKAN WORKSHOP BW2003 *"Mathematical, Theoretical and Phenomenological Challenges Beyond the Standard Model: Perspectives of Balkans Collaboration"* (Vrnjacka Banja, Serbia, August 29 - September 3, 2003, http://www.pmf.ni.ac.yu/bw2003) came to a common agreement on the Initiative for the SEENET-MTP NETWORK. Within the years 2004-2009, 16 institutions from 7 countries (Bosnia and Herzegovina, Bulgaria, Croatia, Greece, Romania, Serbia, Turkey) in the region joined the Network (http://www.seenet-mtp.info/network-nodes.html), 12 partner institutions all over the world (http://www.seenet-mtp.info/partners.html), as well as about 200 individual members (http://www.seenet-mtp.info/members.html), with an increasing trend. This Network has been a natural continuation of Prof. Julius Wess` initiative: Scientists in Global Responsibility (Wissenschaftler in Global Verantwortung – WIGV in German), started in 1999.

Southeastern European Network in Mathematical and Theoretical Physics (SEENET MTP) consists of Faculties (Departments) and research institutes from six Southeastern European countries. The full list of the partner institutions participating in the SEENET MTP is presented in the Appendix A. In accordance with the decisions made at the last meeting of the Representative Committee of the Network (April 2009), leading structure of the Network is as follows

# STRUCTURE AND GOALS

In accordance with the decisions made at the last meeting of the Representative Committee of the Network (April 2009), leading structure of the Network is as follows

*President of the Network*
**Prof. Dr Radu Constantinescu,** Faculty of Physics, University of Craiova, Romania.

*Executive Director of the Network*
**Prof. Dr Goran Djordjevic,** Executive Director of the Network and its permanent Office in Nis, Department of Physics, Faculty of Science, University of Nis, Serbia.

*Executive Committee (EC)*
The primary objectives of the EC are the elaboration of the Network Program, its implementation and widening of the Network financial basis. The EC consists actually of 6 representatives from 7 countries (http://www.seenet-mtp.info/committees/executive-committee). Depends of the budget, the EC meets regularly once per year. Administration of the Network has been realized by the SEENET-MTP Office at the University of Nis - Faculty of Science and Serbian Physical Society-Section Nis. In addition, they are responsible for maintaining and improving the official **web site** (http://www.seenet-mtp.info) of the SEENET-MTP.

*Representative Committee (RC)*
Each of 16 Network nodes is represented by one person (http://www.seenet-mtp.info/representative-committee.html), and included in preparation and implementation of the Network activities. Relation between all Committees is specified by the Terms of references accepted in November `05.

*Scientific-Advisory Committee (SAC)*
The Scientific-Advisory Committee (SAC) comprises a number (actually 17) of outstanding and leading international researchers from both the SEE region and elsewhere in the world (http://www.seenet-mtp.info/committees/scientific-advisory-committee). The main responsibilities of the SAC are: consideration of research project proposals submitted by the nodes of the SEENET-MTP; recommendation on promising research topics of high international interest; support of the SEENET-MTP activities and participation in the SEENET-MTP events; supporting bilateral and multilateral co-operation between the SAC members' home institutions and the SEENET-MTP. The coordinator of the SAC is Prof. Goran Senjanovic from ICTP Trieste.

As *general objectives* of the Netwok one can mention:
- strengthening of close relations and co-operation among faculties of science, research institutions and individual scientists across the region of South Eastern Europe (SEE);
- joint scientific and research activities in the region and improvement of the interregional collaboration through networking, organization of scientific events and mobility programs;
- support capacity building in science and technology for development by initiating intensive, new approaches to teaching physics and sciences, promotion of science through exhibitions, popular lectures, publications, meetings and contests;
- promote exchange of students and encourage communication between gifted pupils motivated for natural sciences and their high schools at regional and interregional level;
- support establishment of local and regional centers of excellence in Sciences.
- to create a database as a basement for a refreshable overview on results obtained by different research organizations and, by that, on the institutional capacity-building in Physics and Mathematics.

# MAIN ACTIVITIES AND RESULTS

Mobility, research, organization of joint scientific meetings, capacity building and promotion of science for youth have been main activities of the Network since its establishing in 2003.

Promotion of science through the Network, in particular physics, started in Serbia in 2003, has been extended to much broader public, covering, in a sense the whole region. Popular lectures adjusted to the target audience had an important scientific and cultural dimension, allowing scientific and students` community in the region to be aware of the scientific capacity and cultural tradition and trends in various countries in the region. In general, promotion of sciences for undergraduate students and gifted high school pupils has been one of the priorities in the whole Network program. The first Romania-Serbia meeting of high school pupils and undergraduate students (http://cis01.central.ucv.ro/physics/ro/info_general/nis.htm), held in Craiova in the autumn 2008, was the first regional meeting of this kind

## Popular lectures and publications

An important part of the activities was the organization of attractive public lectures. The way of dealing with modern topics in physics, cosmology, biology-genetics and related subjects was adjusted to high school students. Almost all of the lecturers and lectures were very attractive and gathered audience of hundreds of people (http://www.seenet-mtp.info/category/news/, http://tesla.pmf.ni.ac.yu/pop-science). The lectures are very useful for teachers as well, and available through Internet or as (two already issued) books with the corresponding CD. One of the most important activities of the SEENET-MTP Network in this field, is the publication of the book on "Modern Topics in Science", basically in Physics, in cooperation with the Serbian Ministry of Science, a bilingual one, English and Serbian. The book consists of popular-nontechnical articles of the leading researchers, members and partners of the Network, suitable for a broader public, especially for the teachers and gifted pupils and young students.

## Exchange and mobility

During last 6 years, about 100 exchanges of researchers and students were completely or partially finances by the Network. It was a tremendous increasing of mobility in the Blakan region, badly suffered for decades because of lack of financing and possibilities for mobility

The list of the scientists included in this part of program is used to be created on the proposals made by the members of the Representative Committee (RC) and coordinated by the members of the Executive Committee (EC). The program was implemented on the basis of scientific and teaching interests of the research group, scientific background of the invited guests, ``quality`` of students and young researchers. In the implementation of the mobility a balance between geographical representation, local financial support and other contributions from the host nodes were considered.

The SEENET-MTP Network, in collaboration with UNESCO-BRESCE Venice and ICTP Trieste provided 3 short-term fellowships (http://www.seenet-mtp.info/breaking-news/short-term-fellowships/), with one month duration, to PhD and MSc students (theoretical, mathematical physics or an adequate) from SEE countries already joined the SEENET-MTP Network (Bulgaria, Croatia, Greece, Romania, Serbia and Turkey). The conditions for the application were that a PhD or MSc student, has a very good record through study, in particular in theoretical and mathematical physics, strong motivation for research, ability to communicate and collaborate, and good knowledge of English. The fellowship provided by the Network was up to 1000 USD, paid through the host institution. The List of the Institutions, groups and individual researchers with the proposed topics to be studied during the visit were announced at http://www.seenet-mtp.info/misc/fellowships/. There were 11 regular applications (before the deadline and with the complete information about the applicants) from Romania, Bosnia and Herzegovina, Croatia and Serbia. The board consisting of 5 members of the SEENET-MTP committees decided to grant Marko Simonovic, Milos Cvetkovic and Marija Marjanovic. They choose to use their grants in Romania (University of Craiova and University of Timisoara). Reports from students and their local advisers show that this first step in mobility inside SEE region was extremely useful and successful.

## Conferences, Schools, Workshops

Being established at the First Balkan Workshop BW2003, the Network, somehow keep organization of the meetings as one of the cornerstone in its ``early`` phase of activities. The Network organized or supported in a good percentage 12 scientific events and supported even more. A comprehensive but quite complete list of the Network meetings could be found at http://www.seenet-mtp.info/category/events/network-meetings/ .

## Research and publications

Closer contacts, exchange and joint conferences with leading researchers and members of the leading groups in Europe in many fields in Theoretical nad Mathematical Physics has as a result numerous join publications and even several projects. We will have note just 3 main publications as a result and review of ``stay of art`` in High Energy Physics, presented through one book and two special issues of the Scientific journals [1-3]. It is evident that a closer focused research needs a bit different financing, not possible through UNESCO and similar Institutions. Because of that, leading structures of the Network are making efforts to establish several working groups and prepare projects for COST nad other similar programs, directly or indirectly related to the FP7 program and other possible funds.

## THE NETWORK AND CHALENGES IN HIGHER EDUCATION

In the next future the efforts of the leading structures of the network will be directed towards the identification of the groups of excellence in Physics and Mathematics in South East Europe and towards strengthening of the interest for these disciplines, especially for youth and women. Two groups of activities devoted to these main objectives are proposed in the form of two distinct sub-projects integrated in this application:
- Map of the top research groups and organizations in the two fields, following their publication records from 2000-2009.
- Promotion of the excellence and growing of the youth's interest for education in Physics and Mathematics.

The first objective will be materialized by an overview on the actual contribution of the countries from South East Europe (SEE) to the world-wide scientific production in Mathematics and Physics. The evaluation of the results in science represents an important concern at the international level. Many studies tried to establish the most adequate criteria and strategies [1]. There is an extended literature about the possibility to combine different indicators in order to obtain a realistic image on the status of the scientific and technological excellence. In the last decade, almost all authors agreed on the fact that the "bibliometric" observation of publications in scientific journals or of the quotations in the major bibliographical databases offers the best image on the level of scientific production and collaboration [2]. This consensus comes from the natural assumption that scientists who have to say something important do share their findings with the others, by publishing them into international journals. This is why, a lot of analysis, applying various quantitative techniques (statistics, data analysis, etc.) have been done using the international databases.

The results of our study could be very useful in the elaboration of a national strategy of scientific development in each SEE country, and of a scientific cooperation strategy in the region. The survey will take into consideration the following countries: Albanie, Bosnia and Herzegovina, Bulgaria, Croatia, Cyprius, Macedonia, Montenegro, Romania, Serbia , Turkey. The research should last for 18 months, from Nov.2009 to April 2011.

As far as the promotion of the excellence and growing of the youth's interest for education in Physics and Mathematics, we aim to contribute to the cooperation of scientists in the region and training programs for graduate students and young researchers. These activities will also have an important impact to the promotion the role of science in the contemporary society and growing of the youth's interest for education and working in Sciences, in particular in Physics and Mathematics. Two main activities will be implemented in the frame of this objective:

**A. The 7-th Spring School and Workshop in Quantum Field Theory and Hamiltonian Systems**
- Date and place: 6-12 May 2010, Calimanesti, Romania.
- Participants: More than 50 students and researchers from the region, as well as 10 outstanding specialists allover the world, who will give lectures.

**B. Exchange and mobility Program.** We intend to continue the tradition of organizing exchanges between scientists and student mobility in the region.

## EXPECTED RESULTS AND OUTCOMES

Although there are few programs that sustain contacts and mobility of researchers and young studious people from the Balkan states to the Western part of Europe, there is lack of opportunities that would make possible for young people to get acquainted with their regional neighborhood. The project we propose could allow for establishing a self-sustained Balkan Research and Education Network in Physics and Mathematics preferably focused on a few fields of research through the Network groups. The support that we will obtain could help to reinforce and extend the previous activities, by enlarging the number of active researchers, graduate and undergraduate students. The proposed meetings, mobility, scientific talks and the Network web portal would have an important scientific and cultural dimension, allowing scientific and student's community in the region to be aware of the scientific capacity and cultural tradition and trends in various countries in the region. As concrete expected results of this project we mention:

- A report concerning the ranking of the universities and research institutes in the SEE region will be elaborated. Based on this report, strategies of involving these institutes in international scientific co-operations can be formulated. The study will also offer an increased visibility of the organizations from the region with the best result in Physics and Mathematics. The report will be an intermediary one, the final report asking for a supplementary validation which could be done in the frame of a forthcoming project.

- We expect that the proposed Network meetings and exchanges will be extremely useful in continuation of scientific exchange of new ideas and results between researchers from the region and top level researchers from Europe and other parts of the world. These meetings will be just partially supported from the UNESCO-Network funds.

## CONCLUSION

Establishing of the specialized working groups is still one of the highest priorities. It is expected, and already confirmed at least in one case, that working groups will be able to apply and get additional support from funds out of UNESCO field of activities in the near future. Hosting and maintaining of the SEENET-MTP web server has been of GREAT importance for keeping contact between the Network nodes and members. In the same time it grows, being a very useful information source for many researchers (containing meeting information including talks presentation, colloquiums, invited lectures and seminars, schools, grants, technical support for numerical computation etc) with fast increasing number of ``hits``. Promotion of sciences for undergraduate students and gifted high school pupils has been one of priorities in the global Network program. An original program related to the Special class for gifted pupils in science in Nis has already attracted more than one thousand of pupils to: the public lectures given by guest lecturers included in the Network program, the educational excursion and competitions. A corresponding book containing ten lectures with multimedia CD has been published. The two more publications are in preparation, with a clear indication of the UNESCO – SEENET-MTP support. It is expected that this part of the project will strengthen similar activities in the all Network nodes and support just established contacts between similar school classes in Serbia, Bulgaria, Romania and others.

# ACKNOWLEDGMENTS

Southeastern European Network in Mathematical and Theoretical Physics was created and permanently supported by UNESCO Office in Venice. In particular, one can mention UNESCO-ROSTE grants No. 8759145, and UNESCO-BRESCE grants No. 8758346 and 8759228. Substantial support comes through UNESCO International Basic Science Program (IBSP) and its funds. There has been a continuous support from ICTP Trieste, Serbian Ministry for science and several Romanian institutions, in particular University of Craiova. We kindly acknowledge to them and many other supporters of the Network and Network`s activities

We would like to underline crucial support of Prof. Julius Wess (and his group at LMU Munich) from 2000 to his untimely death in 2007.

# REFERENCES

1. *EUROPEAN COMMISSION, Key Figures 2007, Towards a USDopean Research Area Science, Support for the coherent development of policies, Directorate-General for Research, EUR 22572 EN*.
2. *INDICATUSDS DE SCIENCES ET DE TECHNOLOGIES, Édition 2008, OST rapport, Éditions Economica & OST*.
3. *Mathematical, Theoretical And Phenomenological Challenges Beyond The Standard Model,* edited by G. S. Djordjevic, Lj. Nesic and J. Wess, World Scientific, Singapore, 2005.
4. *FactaUniversitatis, SeriesPhysics, Chemistry and Technology,* Vol **4**, No 2, 133-449 (2006).
5. *Fortschritte der Physik*, **4-5,** 305-551 (2008)